\documentstyle[12pt]{article}
\begin{document}
\noindent
\begin{center}
{\Large {\bf Mach's Principle and Model for a Broken Symmetric
Theory of Gravity\\}} \vspace{2cm} ${\bf Yousef~Bisabr}$
\footnote{e-mail:~y-bisabr@srttu.edu.}\\
\vspace{0.5cm} {\small {Department of Physics, Shahid Rajaee
University, Lavizan, Tehran 16788, Iran.}}

\end{center}
\vspace{1cm}
\begin{abstract}
We investigate spontaneous symmetry breaking in a conformally
invariant gravitational model. In particular, we use a
conformally invariant scalar tensor theory as the vacuum sector of
a gravitational model to examine the idea that gravitational
coupling may be the result of a spontaneous symmetry breaking. In
this model matter is taken to be coupled with a metric which is
different but conformally related to the metric appearing
explicitly in the vacuum sector. We show that after the
spontaneous symmetry breaking the resulting theory is consistent
with Mach's principle in the sense that inertial masses of
particles have variable configurations in a cosmological context.
Moreover, our analysis allows to construct a mechanism in which
the resulting large vacuum energy density relaxes during
evolution of the universe.
\end{abstract}
\vspace{4cm}
\section{Introduction}
One of the interesting possibilities concerning the origin of
gravitational coupling is that it may be the result of an
invariance breaking of some fundamental symmetry of nature. In
recognition of such a symmetry, the first possibility may be that
like other fundamental interactions, the coupling constant of
gravitational interaction has its origin in a spontaneous
symmetry breaking at some appropriately energy scales.  In fact
Zee \cite{zee} has employed a scalar tensor theory to show that a
spontaneous symmetry breakdown at Planck scale would lead to a
gravitational coupling as suggested by general relativity. In
this approach one recognizes two problems: firstly, the model
seems not to be consistent with Mach's principle in the sense
that after the symmetry breaking the model reduces to general
relativity and specifically the gravitational coupling is given
by the gravitational constant\footnote{It should be remarked that
Mach's idea on the nature of inertia has found a limited
expression in general relativity.  For a detailed discussion see,
for example, \cite{mach}.}. The implication is that scalar tensor
theories motivated by Mach's principle \cite{mach} have no
relevance at energy scales lower than the Planck scale. Secondly,
it is well known that the proposed symmetry breaking leads to the
appearance of a vacuum energy density which is enormously larger
than the experimental upper limit \cite{wein}.\\
There is another possibility about the kind of symmetry which may
be of significance.  Since gravitational coupling is a dimensional
coupling, its strength can be changed by a unit transformation.
Thus the corresponding symmetry which is expected to have
important role is conformal symmetry.  One may study an
invariance breaking effect in a conformally invariant
gravitational model by introducing a constant mass scale.  It has
been shown \cite{deser} \cite{sb} that such an invariance breaking
also leads to a gravitational coupling with the same strength as
used in general relativity.\\
Our main purpose in the present work is to keep the idea that the
gravitational coupling arises from an invariance breaking effect
with special concern for addressing the two aforementioned
problems. We shall study spontaneous symmetry breaking in a
conformally invariant scalar tensor theory in which the scalar
field has a quartic self-interaction term.  The basic ingredient
in the theory is that we take the two metric tensors describing
the gravitational and the matter parts to belong to different
conformal frames.  We then consider the conformal factor relating
the two metric tensors as a dynamical field. Such a dynamical
field is basically imposed in our model to make a dynamical
distinction between the two unit systems usually used in
cosmology and elementary particle physics. We have already
emphasized the significant role of this dynamical distinction in
construction of a mechanism to reduce a large cosmological
constant during evolution of the universe \cite{bs}.  In the
present work we intend to investigate the role of conformal
symmetry in a scalar tensor theory that undergoes spontaneous
symmetry breaking. We argue that if the theory is taken to be
conformally invariant, after the spontaneous symmetry breaking it
remains consistent with the spirit of Mach's principle.  We also
show that the resulting vacuum energy density
appears as a decaying cosmological constant.\\
We organize this paper as follows: In section 2, we first offer a
brief review of the model proposed by Zee \cite{zee}. In section
3, we use a gravitational model which is conformally invariant.
We argue that there is an ambiguity concerning the coupling of
matter systems to such a model. In general matter systems should
be coupled to the metric which is conformally related to that
describing the vacuum sector.  In section 4, we consider
spontaneous symmetry breaking in this gravitational model. We
discuss the problem of the emergence of a large vacuum energy
density and consistency of the model with
Mach's principle.  In section 5, we outline our results.\\
Throughout the following we shall
use units in which $\hbar=c=1$ and the signature is (-+++).\\
\section{A brief review of Zee's model}
Spontaneous symmetry breaking is one of the key ideas in
elementary particle physics which is expected to lead to
unification of strong, weak and electromagnetic interactions.  In
order to incorporate this mechanism into gravity, Zee proposed
\cite{zee} a modification of the Einstein-Hilbert action used in
general relativity.  The proposed model is
\begin {equation}
S=-\int d^4 x \sqrt{-g} \{\frac{1}{2} \phi^2
R+\frac{1}{2}g^{\mu\nu}\nabla_{\mu}\phi
\nabla_{\nu}\phi-V(\phi)\}+S_{m}(g_{\mu\nu},\phi)~,
\label{a1}\end{equation} where $\phi$ is a scalar field with a
potential $V(\phi)$, $R$ is the curvature scalar, $\nabla_{\mu}$
denotes covariant differentiation and $S_{m}(g_{\mu\nu},\phi)$ is
the matter field action. In this gravitational system $\phi^{-2}$
characterizes the gravitational coupling.  Moreover, the scalar
field $\phi$ plays the role of the Higgs field and should
therefore have interaction with the matter part of the model so
that in the
above action $S_{m}(g_{\mu\nu},\phi)$ includes $\phi$.\\
variation of the action (\ref{a1}) with respect to $g^{\mu\nu}$
and $\phi$ gives, respectively,
\begin{equation}
\phi^2 G_{\mu\nu}=T_{[m]\mu\nu}+T_{[\phi]\mu\nu}~,
\label{a2}\end{equation}
\begin{equation}
\Box\phi-\phi R+\frac{\partial V}{\partial
\phi}=\frac{1}{\sqrt{-g}}\frac{\delta
S_{m}(g_{\mu\nu},\phi)}{\delta \phi}~, \label{a3}\end{equation}
where
\begin{equation}
T_{[m]\mu\nu}=-\frac{2}{\sqrt{-g}}\frac{\delta
S_{m}(g_{\mu\nu},\phi)}{\delta g^{\mu\nu}}~,
\label{a4}\end{equation} and
\begin{equation}
T_{[\phi]\mu\nu}=-(\nabla_{\mu}
\phi\nabla_{\nu}\phi-\frac{1}{2}g_{\mu\nu}\nabla_{\gamma}\phi\nabla^{\gamma}\phi)
-\frac{1}{2}(g_{\mu\nu}\Box\phi^2-\nabla_{\mu}\nabla_{\nu}\phi^2)-g_{\mu\nu}V(\phi)~.
\label{a5}\end{equation} Here $\Box \equiv
g^{\mu\nu}\nabla_{\mu}\nabla_{\nu}$ and $G_{\mu\nu}$ is the
Einstein tensor.  The potential $V(\phi)$ usually contains a
quartic self-interaction and an imaginary mass term.  This
suffices to give minima to the potential for some nonvanishing
values of $\phi$. If the degenerate vacuum is $\phi=v$ with $v$
being a constant, the spontaneous symmetry breaking then results
in a gravitational coupling with the same strength as $v^{-2}$.
This gives the gravitational constant if one takes the mass scale
$v$ to be of the same order of the Planck mass. This means that
the spontaneous symmetry breaking should take place at Planck
scale. In lower energy scales, however, there is no difference
between this model and general relativity since with $\phi=v$ the
field equations (\ref{a2}) and (\ref{a3}) reduce to the Einstein
field equation, namely,
\begin{equation}
G_{\mu\nu} +v^{-2}V(v)g_{\mu\nu}=v^{-2} T_{[m]\mu\nu}~.
\label{aa1}\end{equation} It is well known that such a symmetry
breaking induces a large vacuum energy density in the
gravitational equations. This problem is not, however, addressed
in the Zee's model since it is assumed that $V(v)=0$.  In the
following, we shall recognize this vacuum energy problem as one
of the two main problems affecting the model described by the
action (\ref{a1}).  The other problem is that this model is not
consistent with Mach's principle since it proposed that gravity
is described by general relativity in all energy scales lower
than the Planck scale \cite{mach}.
\section{The model}
We consider a scalar tensor theory consisting of a real scalar
field $\phi$, described by the action functional\footnote{This
action is identical to the gravitational part of (\ref{a1}), if
$\frac{1}{2}\phi^2 R$ is replaced by $\frac{1}{12}\phi^2 R$ and
$V(\phi)$ is taken to be $-\frac{1}{2}\lambda \phi^4$. }
\begin{equation}
S=-\frac{1}{2}\int d^4x \sqrt{-g} \{g^{\mu\nu}\nabla_{\mu}\phi
\nabla_{\nu}\phi+\frac{1}{6}\phi^2 R+\lambda\phi^4 \}~,
\label{b4}\end{equation} where $\lambda$ is a dimensionless
coupling constant.  This action is invariant under conformal
transformations
\begin{equation}
\bar{g}_{\mu\nu}=e^{2\sigma}g_{\mu\nu}~, \label{b5}\end{equation}
\begin{equation}
\bar{\phi}=e^{-\sigma}\phi~, \label{b6}\end{equation} where
$\sigma$ is a dimensionless spacetime function. When this action
is taken as the vacuum sector of a gravitational model, one
encounters an inherent ambiguity concerning the incorporation of
matter systems. In fact since the vacuum sector is conformally
invariant, it is not possible to make any distinction between two
different conformal frames and all the frames must be considered
as dynamically equivalent. In this situation it is not clear to
which of these conformal frames, or the corresponding metric
tensors, the matter systems should be coupled. To consider the
most general case, we take the matter systems to be coupled with
the metric $\bar{g}_{\mu\nu}$ rather than $g_{\mu\nu}$ which are
conformally related due to (\ref{b5}).  In this way, we take into
account all the dynamical implications of different conformal
frames. We therefore write the action (\ref{b4}) in the form
{\begin{equation} S=-\frac{1}{2}\int d^4x \sqrt{-g}
\{g^{\mu\nu}\nabla_{\mu}\phi \nabla_{\nu}\phi+\lambda
\phi^4+\phi^2(g^{\mu\nu}\nabla_{\mu} \sigma
\nabla_{\nu}\sigma+\frac{1}{6}R) \}+S_{m}(\bar{g}_{\mu\nu},\phi)~,
\label{b7}\end{equation} where $S_{m}(\bar{g}_{\mu\nu},\phi)$ is
the matter action containing some matter field variables,
including the gauge fields of the standard model, coupled to the
metric $\bar{g}_{\mu\nu}$.  We shall consider the case that the
matter action interacts with the scalar field $\phi$. This is
necessary for interpretation of the scalar field as a Higgs
field. We have also taken the above action to involve a kinetic
term for $\sigma$ to account for its dynamical contributions. In
this case the vacuum sector of the action (\ref{b7}) is still
conformally invariant since $\sigma$ as a dimensionless function
does not
change under conformal transformations.\\
Variation of the action (\ref{b7}) with respect to $g^{\mu\nu}$,
$\phi$ and $\sigma$, yields,
\begin{equation}
G_{\mu\nu}-3\lambda \phi^2
g_{\mu\nu}=6\phi^{-2}(T_{\mu\nu}(\bar{g}_{\mu\nu})+\tau_{\mu\nu})+6t_{\mu\nu}~,
\label{b8}\end{equation}
\begin{equation}
\Box \phi-\frac{1}{6}R\phi-2\lambda \phi^3-\phi
\nabla_{\gamma}\sigma \nabla^{\gamma}\sigma=-\frac{\delta}{\delta
\phi }S_{m}(\bar{g}_{\mu\nu},\phi)~, \label{b9}\end{equation}
\begin{equation}
\nabla_{\mu}(\sqrt{-g} \phi^2 g^{\mu\nu}
\nabla_{\nu}\sigma)=\sqrt{-g}g^{\mu\nu}T_{\mu\nu}(\bar{g}_{\mu\nu})~,
\label{b10}\end{equation} where
\begin{equation}
T_{\mu\nu}(\bar{g}_{\mu\nu})=\frac{2}{\sqrt{-g}}\frac{\delta}{\delta
g^{\mu\nu} }S_{m}(\bar{g}_{\mu\nu},\phi)~,
\label{b11}\end{equation} and
\begin{equation}
\tau_{\mu\nu}=-(\nabla_{\mu}\phi
\nabla_{\nu}\phi-\frac{1}{2}g_{\mu\nu}\nabla_{\gamma}\phi
\nabla^{\gamma}\phi)-\frac{1}{6}(g_{\mu\nu}\Box-\nabla_{\mu}
\nabla_{\nu})\phi^2~, \label{b12}\end{equation}
\begin{equation}
t_{\mu\nu}=-(\nabla_{\mu} \sigma
\nabla_{\nu}\sigma-\frac{1}{2}g_{\mu\nu} \nabla_{\gamma} \sigma
\nabla^{\gamma}\sigma)~. \label{b13}\end{equation}
We would like to consider breakdown of the conformal invariance
in the action (\ref{b7}). Therefore we add a term such as
\begin{equation} -\frac{1}{2}\int
d^{4}x \sqrt{-g}m^2 \phi^2~, \label{bb15}\end{equation} to this
action with $m$ being a constant mass scale. In this case
equations (\ref{b8}) and (\ref{b9}) change to
\begin{equation}
G_{\mu\nu}-3(m^2+\lambda \phi^2)
g_{\mu\nu}=6\phi^{-2}(T_{\mu\nu}(\bar{g}_{\mu\nu})+\tau_{\mu\nu})+6t_{\mu\nu}~,
\label{17}\end{equation}
\begin{equation}
\Box \phi-\frac{1}{6}R \phi-m^2 \phi-2\lambda \phi^3 -\phi
\nabla_{\gamma}\sigma \nabla^{\gamma}\sigma=-\frac{\delta}{\delta
\phi}S_{m}(\bar{g}_{\mu\nu},\phi)~, \label{18}\end{equation}
Looking at the gravitational equation (\ref{17}) one infers from
the sign of the mass term that it has a negative contribution to
vacuum energy density.  This feature seems to be generic to all
scalar tensor theories (namely theories that consider a scalar
field with nonminimal coupling to gravity) which entail a massive
scalar field. The sign of this mass term can however be changed by
introducing a constant mass scale such as $\mu$ with
$\mu^2=-m^2$. It is important that this induces spontaneous
symmetry breaking in the action (\ref{b7})\footnote{Note that
$\mu$ appears as a tachyonic mass in the field equation
(\ref{18}).}. In this case the potential of the scalar field can
be written as $U(\phi)=-\mu^2 \phi^2+\lambda\phi^4$. The minimum
of this potential is determined by the conditions
\begin{equation}
\frac{d U}{d\phi}=-2\mu^2 \phi+4\lambda\phi^3=0~,
\label{c2}\end{equation} and
\begin{equation}
\frac{d^2 U}{d \phi^2}=-2\mu^2+12\lambda\phi^2>0~,
\end{equation}
where $\lambda>0$.  The relation (\ref{c2}) has nonzero solutions
$\phi_{0}^2=\frac{\mu^2}{2\lambda}$ minimizing the potential at
$U(\phi_{0})=\frac{-\mu^4}{4\lambda}$. The gravitational coupling
is then given by $\phi^{-2}_{0}\sim \lambda \mu^{-2}$. This is the
gravitational constant if $\mu \sim \lambda^{\frac{1}{2}} m_{p}$
with $m_{p} \sim G^{-\frac{1}{2}}$ being the Planck mass. Note
that the energy scale at which the spontaneous symmetry breaking
takes place is not necessarily the Planck scale.  It is given by
$\mu$
which depends on the precise value of the coupling constant $\lambda$. \\
For $\phi=\phi_{0}$, the action (\ref{b7}), together with
(\ref{bb15}) and $\mu^2=-m^2$, reduces to {\begin{equation}
S=-\frac{1}{16\pi G}\int d^4x \sqrt{-g}
\{R-2\Lambda+6g^{\mu\nu}\nabla_{\mu} \sigma \nabla_{\nu}\sigma
\}+S_{m}(\bar{g}_{\mu\nu})~, \label{c3}\end{equation} where
$\Lambda=\frac{3}{2}\mu^2$. It is clear that the result of the
spontaneous symmetry breaking in (\ref{b7}), namely the action
(\ref{c3}), is very different from that obtained by the action
(\ref{a1}).  In the action (\ref{c3}), the metric tensors in the
gravitational and the matter parts belong to different conformal
frames and the conformal factor itself appears as a dynamical
field. The consequences of such a coupling are discussed in the
next section.
\section{Mach's principle and the cosmological constant}
When a constant mass scale such as $m^2$ (or $-\mu^2$) is
introduced into the gravitational part of the action (\ref{b7}),
the conformal invariance is broken and a particular conformal
frame (or unit system) is singled out in terms of the values
attributed to the dimensional quantities. In general, the use of
two different unit systems are conventional. In one hand, the
gravitational constant as a dimensional coupling characterizes a
unit system usually used in cosmology which is referred to as the
cosmological frame \cite{bs}.  On the other hand, there is a
particle unit system used in elementary particle physics and is
defined in terms of compton wavelenght of a typical elementary
particle.  It is important to note that one usually assumes that
these two unit systems are related by a global unit
transformation. This means that they are taken to be
indistinguishable up to a constant conversion factor in all
spacetime points.  Such a unit transformation is obviously devoid
of any dynamical implication. Here we consider a local unit
transformation \cite{dick} which requires that the two different
unit systems be interrelated by a spacetime dependent conversion
(or conformal) factor. This gives a dynamical meaning to changes
of unit systems and seems to be more consistent with Mach's
principle. The reason for this is
discussed in the following.\\
According to Mach's principle, one may attribute inertial forces
experienced in a given point of spacetime to the gravitational
forces due to distant accelerated matter systems.  By
implication, the inertial mass of a particle should depend on the
distribution of matter around that particle.  As a consequence one
expects that inertial masses have different values in different
spacetime points.\\
On the other hand, there is an inherent ambiguity concerning
measurement of changes of a dimensional quantity.  In general, the
value of a dimensional quantity not only may change in a given
unit system but it may also change due to changes of the unit
system by which the quantity is measured. There is not however any
direct way to distinguish between these two types of changes.
Thus, it is only meaningful to compare mass ratios, as
dimensionless quantities, rather than mass itself at different
spacetime points.  For construction of such a dimensionless
quantity one may take proportion of the inertial mass of a
typical elementary particle, $M$, and the Planck mass, namely,
\begin{equation}
 M(G)^{\frac{1}{2}}=n~,
\label{d1}\end{equation} where $n$ is a dimensionless number.  If
one takes $n$ to vary, as suggested by Mach's principle, one may
consider two possibilities depending on attribution of these
variations to either $M$ or $G$. In the particle frame one takes
$M$ as a constant and $G$ as varying while in the cosmological
frame inertial masses of elementary particles are taken to be
varying and $G$ is regarded as a constant. It is now clear from
the discussion that these two unit systems should be taken to be
related by a spacetime dependent conversion factor in order to
respect the precise statement of Mach's principle.  It should be
remarked that from a physical point of view there seems to be no
fundamental difference between the two possibilities although
their precise formulations may need theories which have quite
different mathematical structures.  In the present work, we shall
not concern with this issue and only the physical content of the
dynamical distinction of the two unit systems is brought into
focus.\\
Now we turn to interpret the action (\ref{c3}). For doing this, we
recall that when the action (\ref{a1}) undergoes spontaneous
symmetry breaking both the gravitational coupling and the
inertial masses have constant configurations contrary to the
statement of Mach's principle. On the other hand, the appearance
of $\sigma$ as a dynamical field in the action (\ref{c3}) implies
that inertial masses of elementary particles have variable
configurations even though the gravitational coupling takes a
constant value\footnote{It is important to note that when the
conformal invariance of the gravitational part of (\ref{b7}) is
broken the resulting theory automatically chooses the
cosmological frame. This is a direct consequence of the
spontaneous symmetry breaking in the action (\ref{b7}) that gives
a constant configuration to the scalar field.}. This is due to
the fact that matter is coupled to gravity through the metric
$\bar{g}_{\mu\nu}$ which is conformally related to $g_{\mu\nu}$.
We emphasize that this feature is a direct consequence of the
conformal invariance of the vacuum sector of the action (\ref{b7})
which gives plausibility to consider such a coupling.\\
As the last point, we investigate the appearance of $\Lambda$ in
(\ref{c3}) as a large effective vacuum energy density.  We remark
that this does not lead to the cosmological constant problem in
the context of our model which assumes that the cosmological and
the particle unit systems are dynamically distinct. To clarify
this point we first take the two metric tensors $g_{\mu\nu}$ and
$\bar{g}_{\mu\nu}$ to describe the cosmological and the particle
frames, respectively. They are related by
$g_{\mu\nu}=e^{-2\sigma}\bar{g}_{\mu\nu}$. Such a distinction
should also be imposed on the dimensional quantities $\Lambda$
and $\bar{\Lambda}$, namely the value of vacuum energy density in
the cosmological and the particle unit systems. According to the
dimension of $\Lambda$ (the squared mass) this two quantities are
related by $\Lambda=e^{2\sigma}\bar{\Lambda}$. Thus $\Lambda$ is
not actually a constant in the cosmological frame and the action
(\ref{c3}) should be written as {\begin{equation}
S=-\frac{1}{16\pi G}\int d^4x \sqrt{-g}
\{R-2\bar{\Lambda}e^{2\sigma}+6g^{\mu\nu}\nabla_{\mu} \sigma
\nabla_{\nu}\sigma \}+S_{m}(\bar{g}_{\mu\nu})~,
\end{equation} which leads to the field equations
\begin{equation}
G_{\mu\nu}+\bar{\Lambda} e^{2\sigma}g_{\mu\nu} =8\pi G
T_{\mu\nu}(\bar{g}_{\mu\nu})+6t_{\mu\nu}~,
\label{d2}\end{equation}
\begin{equation}
\Box \sigma+\frac{1}{3}\bar{\Lambda}e^{2\sigma}=\frac{4\pi}{3} G
g^{\mu\nu} T_{\mu\nu}(\bar{g}_{\mu\nu})~. \label{d3}\end{equation}
In these equations the exponential coefficient for
$\bar{\Lambda}$ emphasizes the dynamical distinction between the
cosmological and the particle unit systems.  One intuitively
expects that this distinction be indistinguishable immediately
after the spontaneous symmetry breaking. This means that
$g_{\mu\nu}$ and $\Lambda$ coincide with their corresponding
quantities in the particle unit system, namely $\bar{g}_{\mu\nu}$
and $\bar{\Lambda}$, at sufficiently early times.  When the
universe expands, cosmological scales enlarge and this
distinction increases so that $e^{-2\sigma}$ must be an
increasing function of time in an expanding universe. It follows
that $\Lambda$ characterizing the effective cosmological constant
in the cosmological frame damps due to the cosmic expansion
\cite{bs}.\\ As an illustration, we first combine (\ref{d3}) with
the trace of (\ref{d2}) that gives
\begin{equation}
\Box \sigma
+\nabla_{\gamma}\sigma
\nabla^{\gamma}\sigma+\frac{1}{6}R-\frac{1}{3}\bar{\Lambda}e^{2\sigma}=0~.
\label{d4}\end{equation} If we write this equation in a spatially
flat Friedmann-Robertson-Walker spacetime, we obtain\footnote{Due
to homogeneity and isotropy of the universe $\sigma$ is taken to
be only a function of time.}
\begin{equation}
\ddot{\sigma}+3\frac{\dot{a}}{a}\dot{\sigma}+\dot{\sigma}^2-\frac{\ddot{a}}{a}
-\frac{\dot{a}^2}{a^2}+\frac{1}{3}\bar{\Lambda}e^{2\sigma}=0~,
\label{d5}\end{equation} where $a(t)$ is the scale factor and the
overdot denotes
differentiation with respect to time.\\
Now assuming that the universe follows a power-law expansion,
namely that $\frac{\dot{a}}{a} \sim t^{-1}$, we use the ansatz
\begin{equation}
e^{-\sigma}=\sigma_{0}t~, \label{d6}\end{equation} which
satisfies equation (\ref{d5}) with $\sigma_{0}\sim
\sqrt{\bar{\Lambda}}$. The vacuum energy density in the
cosmological frame is then $\Lambda=\bar{\Lambda}e^{2\sigma}\sim
t^{-2}$, consistent with the observational upper limit.
\section{Concluding remarks}
We have investigated the idea that gravitational coupling may be
the result of a mechanism involving the breakdown of some
fundamental symmetry of nature.  Two kinds of symmetry breaking
seem to be relevant on the subject: firstly, it is natural to
think about spontaneous symmetry breaking analogous to the origin
of coupling constants corresponding to other fundamental
interactions and secondly, a conformal symmetry breaking
motivated by the fact
that the gravitational coupling is a dimensional coupling.\\
We discussed spontaneous symmetry breaking in a conformally
invariant scalar tensor theory, in which the scalar field has a
quartic self-interaction, by introducing a constant tachyonic
mass scale. We have shown that a spontaneous symmetry breaking in
the model would lead to a gravitational coupling as suggested by
general relativity.  It should be remarked that this symmetry
breaking may take place in energy scales much lower than the
Planck scale
when $\lambda <<1$.\\
We emphasize that our analysis allows to avoid the two problems
afflict the model proposed by Zee, the action (\ref{a1}), namely
inconsistency with Mach's principle and the cosmological constant
problem.  This is basically due to the fact that in the action
\cite{b7} we couple the matter to a metric which is conformally
related to that describing the gravitational part. In our
approach such a coupling is motivated by the fact that the
gravitational part is conformally invariant and does not
dynamically distinguish between different metric tensors related
by the conformal transformation (\ref{b5}).  The important
feature of this coupling is that it gives variable configurations
to all the mass scales introduced by elementary particle physics
in the cosmological frame. In particular, contributions of these
mass scales in vacuum energy density are
so that they decrease during evolution of the universe.\\
We point out that the nontrivial behaviour of the conformal factor
in our model may have important role in cosmology. For instance,
it may act as a quintessence describing the acceleration of the
universe in the present epoch \cite{acc}.


\newpage



\begin{thebibliography}{99}
\bibitem{zee} A. Zee, Phys. Rev. Lett. {\bf 42}, 417 (1979)
\bibitem{mach} C. Brans and R. H. Dicke, Phys. Rev. {\bf 124}, 925
(1961)
\bibitem{wein} S. Weinberg, Rev. Mod. Phys. {\bf 61}, 1 (1989)
\bibitem{deser} S. Deser, Ann. Phys. {\bf 59}, 248 (1970)\\
H. Salehi, Int. J. Theor. Phys. {\bf 37}, 1253 (1998)
\bibitem{sb} H. Salehi and Y. Bisabr, Int. J. Theor. Phys. {\bf 39}, 1241
(2000)
\bibitem{bs} Y. Bisabr and H. Salehi, Class. Quantum Grav. {\bf
19}, 2369 (2002)
\bibitem{dick} R. H. Dicke, Phys. Rev. {\bf 125}, 2163 (1962)\\
J. D. Bekenstein and A. Meisels, Phys. Rev. D {\bf 22}, 1313
(1980)
\bibitem{acc} Y. Bisabr,  archive no. hep-th/0306192.\\
To appear in International Journal of Theoretical Physics.




\end{thebibliography}
\end{document}